\documentclass[9pt]{IEEEtran}
\pdfoutput=1

\usepackage{color}
\usepackage{spconf}
\usepackage{times}

\usepackage{color}
\usepackage{theorem}
\usepackage{amsmath}
\usepackage{amssymb}
\usepackage{bm}
\usepackage{subfigure}
\usepackage{cite}
\usepackage{hyperref}

\usepackage[font=small,format=plain,up,up]{caption}
\usepackage{graphicx}
\usepackage{needspace}

\input{mysymbol.sty}

\newcommand{\bbeps}{{\boldsymbol{\epsilon}}}

\newtheorem{myproposition}{\bf Proposition}

\newcommand{\QED}{\hfill\ensuremath{\blacksquare}}

\title{Flow Smoothing and Denoising: Graph Signal Processing in the Edge-Space}

\name{Michael T. Schaub$^{a,b}$ and Santiago Segarra$^{a}$
	\thanks{MTS received funding from the European Union's Horizon 2020 research and innovation programme under the Marie Sklodowska-Curie grant agreement No 702410. SS received an IDSS MIT seed grant. The funders had no role in the design of this study; the results presented here reflect solely the authors' views.  Emails: \{mschaub, segarra\}@mit.edu}}
    \address{$^{a}$Institute for Data, Systems, and Society, Massachusetts Institute of Technology, Cambridge, MA, USA\\
    $^{b}$Department of Engineering Science, University of Oxford, Oxford, UK}
\begin{document}
\maketitle
\begin{abstract}
    This paper focuses on devising graph signal processing tools for the treatment of data defined on the edges of a graph. 
    We first show that conventional tools from graph signal processing may not be suitable for the analysis of such signals.
    More specifically, we discuss how the underlying notion of a `smooth signal' inherited from (the typically considered variants of) the graph Laplacian are not suitable when dealing with edge signals that encode a notion of flow.
    To overcome this limitation we introduce a class of filters based on the Edge-Laplacian, a special case of the Hodge-Laplacian for simplicial complexes of order one.
    We demonstrate how this Edge-Laplacian leads to low-pass filters that enforce (approximate) flow-conservation in the processed signals.
    Moreover, we show how these new filters can be combined with more classical Laplacian-based processing methods on the line-graph. 
    Finally, we illustrate the developed tools by denoising synthetic traffic flows on the London street network.
\end{abstract}

\section{Introduction}\label{S:Introduction}

As the availability of relational data continues to increase, graph-based techniques to model, filter, and process such data have become a mainstay in the current literature, cutting across disciplines.
Consequentially, graph signal processing (GSP), which aims to provide a theoretical grounding for the processing of signals defined on graphs, has become a rapidly growing area of research~\cite{SandryMouraSPG_TSP13, shuman2013}.
However, while significant research activity has been devoted to the development of foundational tools for signal processing on graphs, including algorithms for sampling~\cite{marques_sampling_2016, chen_sampling_2015, tsitsvero_sampling_2016}, reconstruction~\cite{segarra_reconstruction_2016, wang_reconstruction_2015, romero_reconstruction_2017}, and filter design~\cite{segarra_filters_2017, isufi_filters_2017, teke_filters_2017}, most research to date has focused on signals defined on the nodes of the graph.

However, in many problems modeled using graphs, the data of interest is located on the edges (as opposed to the nodes) of these graphs.
A typical scenario of practical interest is a flow on the edges -- signal, mass, energy, information -- of a graph that is measured and has to be analyzed further, such as traffic flow associated with the edges of a traffic network~\cite{deri_taxi_2015}.
Further prototypical examples include data emerging from spatially embedded networks such as supply networks (e.g., power grids)~\cite{jablonski_power_2017}, mobility and migration data~\cite{carlsson_migration_2018}, or information flows in brain networks or other biological tissue~\cite{huang_brain_2018,medaglia_brain_2017,schaub_network_traffic_2014,Bacik2016,Amor2016}.
While spatially embedded networks provide many of the most intuitive examples, edge-flows are also the natural objects of interest for web-traffic, click-stream data, and various other communication patterns that are not embedded in any type of physical space.

Despite this large space of applications, there has been comparably little activity related to edge-based analysis of signals on graphs.
One reason for this lack of research activity may be that such situations appear to be transformable into a vertex-based problem by using a line-graph transformation~\cite{godsil_algebraic_2001}.
Line-graphs, which record adjacency relationships between edges, provide indeed some potential modeling strategy to consider edge data.
For instance, line-graphs have been considered for the task of detecting overlapping communities, by clustering edges~\cite{evans_line_graphs_2009,Ahn2010}.

However, as we discuss in this paper, line-graphs are not sufficient as modeling tools for edge data that comes in the form of flows.
As we will see, one deeper underlying reason for this fact is that certain notions from `classical', vertex-based GSP do not seamlessly extend from vertex to edge data:
unlike vertex data, flows typically carry an \emph{orientation}, and accounting appropriately for such orientations can be crucial to reach the desired results.
Interestingly, the edge-space processing perspective that we advocate here may be interpreted in the sense of processing `higher-order' data~\cite{barbarossa_higher_2016, barbarossa_higher_2018,Benson2018}, and can be generalized further by calling upon tools from algebraic topology~\cite{Hatcher2002} and discrete calculus~\cite{Grady2010}, thereby opening up a range of interesting avenues for future research.
However, to motivate our work and provide a concrete problem context, in this paper we focus on the denoising of edge-flow data as a specific example, leaving data defined in higher-order simplicial complexes as a matter of future research.

\noindent \textbf{Contributions and paper outline.} 
We introduce the problem of flow denoising and smoothing for edge-space data.
Specifically, we discuss how the space of edge-flows decomposes into cyclic (harmonic) and gradient flows, and how this leads to a novel notion of low-pass filter for oriented edge-data.
Our proposed filters are based on the Edge-Laplacian $\bbL_1$, a particular form of the more general Hodge-Laplacian encountered in algebraic topology~\cite{Lim2015}.
We show how the Edge-Laplacian $\bbL_1$ solves an optimal filtering problem that aims to filter out gradient flows in the edge-space, and contrast it with the more typically encountered graph Laplacian $\bbL$.

We start in Section~\ref{S:graphs_background} by recalling fundamental notions from GSP.
In Section~\ref{S:motivation_contribution} we first provide an example that motivates the need for edge-space filtering, before discussing some of its various properties more formally in Section~\ref{S:Hodge_filters}.
We finally demonstrate the utility of the developed methodology by denoising traffic flows in the London street network (Section~\ref{S:Simulations}), before concluding with a brief discussion on avenues for future research.

\begin{figure*}[tb!]
    \centering
    \includegraphics[width=0.9\linewidth]{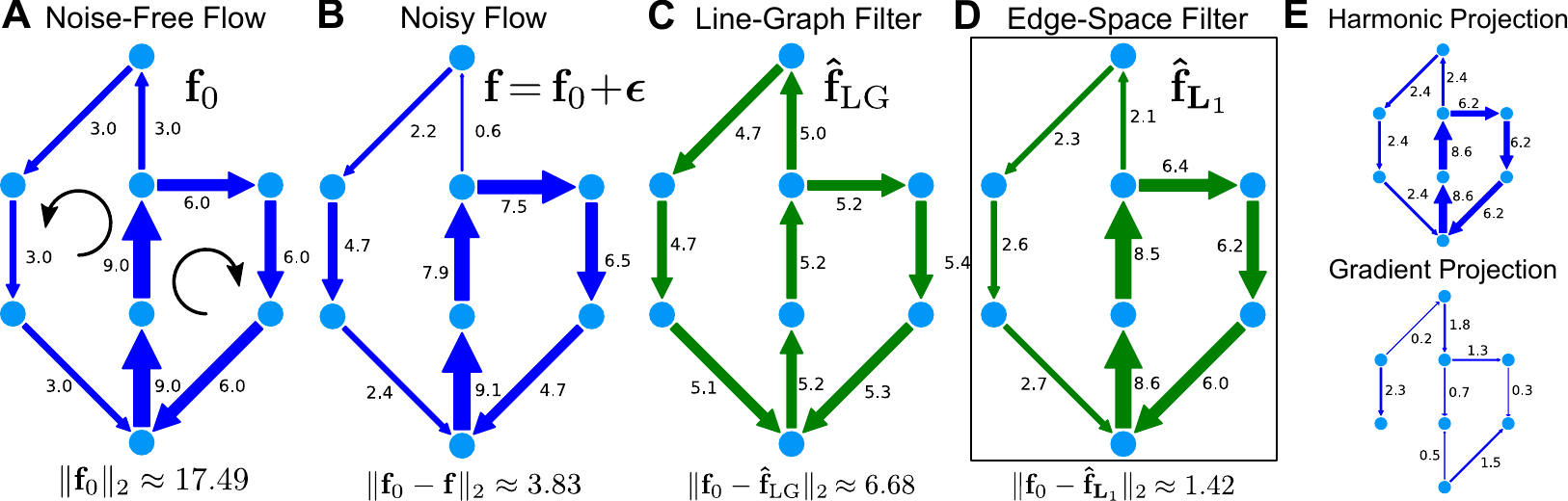}
    \vspace{0.2cm}
    \caption{\textbf{Flow smoothing on a graph -- illustrative example.}
    \textbf{A}~An undirected graph with a pre-defined, oriented flow pattern $\mathbf{f}_0$ on the edges. \textbf{B}~The observation of the flow $\mathbf{f}_0$ is distorted by a Gaussian white noise vector $\bm \epsilon$.
    \textbf{C-D} We may want to denoise the flow by applying a Laplacian filter based on the line-graph of this topology (\textbf{C}). 
However, this does not yields a satisfactory performance, in particular compared to the edge-space filter  approach (\textbf{D}) introduced in this paper. \textbf{E} The originally observed flow $\bbf$ can be decomposed into its harmonic (cyclic) component and its gradient part. Notice that the edge-space filter acts by reducing the gradient flow in the input edge-data.}
    \label{fig:schematic_example}
\end{figure*}

\section{Background}\label{S:graphs_background}
\noindent\textbf{Graphs, incidence matrices, and the graph Laplacian.} 
An~undirected graph $\ccalG$ consists of a node set $\ccalN$ of cardinality $N$ and an edge set $\ccalE$ of unordered pairs of elements in $\ccalN$ of cardinality $E$.
The edges can be conveniently collected as entries of the symmetric adjacency matrix $\bbA$, such that $A_{ij}=A_{ji} = 1$ for all $(i,j)\in\ccalE$, and $A_{ij} = 0$ otherwise.
Defining the diagonal degree matrix $\bbD:=\diag(\bbA\bbone)$, the (combinatorial) Laplacian matrix associated with $\ccalG$ is given by $\bbL:=\bbD-\bbA$, and has eigendecomposition $\bbL = \bbV \bbLambda \bbV^\top$ where the eigenvalues in the diagonal matrix $\bbLambda$ are sorted in increasing order.

An alternative way to encode a graph is by means of a node-to-edge incidence matrix $\bbB \in \reals^{N \times E}$.
To this end, we endow each edge $e$ with an arbitrary reference orientation from its tail node $t(e)$ to its head node $h(e)$. 
The incidence matrix is then defined as ${\bbB_{e\,t(e)} = -1}$, ${\bbB_{e\,h(e)}=1}$ and $\bbB_{ek} = 0$ otherwise.
Based on this definition, it can be shown that the graph Laplacian can be alternatively written as $\bbL = \bbB\bbB^\top$.

\noindent\textbf{Laplacian smoothing and denoising on graphs.}
Consider the observation of a noisy signal vector ${\bby = \bby_0 + \bbeps \in \reals^N}$ defined on the nodes of a connected graph $\ccalG$, where $\bby_0$ is a certain signal of interest and $\bbeps$ represents zero-mean additive noise.
A canonical task in GSP is to provide a filtered signal $\hat{\bby} \approx \bby_0$, based on the observation of the noisy~$\bby$.

A typical assumption made in this context is that the signal should be smooth with respect to the underlying graph.
To leverage this information we can define the following denoising optimization problem 
\begin{equation}\label{E:regularized_least_squares}
\min_\hby \lbrace \|\hby - \bby\|_2^2 + \alpha \hby^\top \bbL \hby \rbrace,
\end{equation}
where $\alpha >0$ can be understood as a regularization parameter that regulates the influence of the smoothness promoting regularizer $\hby^\top\bbL\hby = \sum_{ij}A_{ij}(\hat{y}_i-\hat{y}_j)^2$. 
Notice that this quadratic form can be interpreted as a measure of the variation of the signal on the graph (measured along the edges) and is minimized by a constant vector ${\hby \propto \bbone}$.
The optimal solution of the above optimization problem is given by
\begin{equation}\label{E:node_denoising}
\hby = (\bbI + \alpha\bbL)^{-1}\bby \qquad \text{(node denoising)}. 
\end{equation}
Instead of solving the above problem, another strategy to obtain a filtered signal $\hby$ is to use a simpler iterative smoothing operation for a suitably chosen update parameter $\mu$ and a certain fixed number of rounds $k$
\begin{equation}\label{E:node_smoothing}
    \hby = (\bbI - \mu \bbL)^k \bby \qquad \text{(node smoothing)}.
\end{equation}
Both the denoising and smoothing procedures respectively defined in \eqref{E:node_denoising} and \eqref{E:node_smoothing} are specific instances of \emph{low-pass graph filters}. A graph filter $\bbH$ is a linear map between graph signals that can be expressed as a matrix polynomial of $\bbL$.
A graph filter $\bbH$ is low-pass if its frequency response $\tilde{\bbh} = \diag(\bbV^\top \bbH \bbV)$ is a vector of decreasing values. It is immediate to corroborate that the filters in \eqref{E:node_denoising} and \eqref{E:node_smoothing} are indeed low-pass filters.

\noindent\textbf{Line-graphs and their algebraic representation.}
The line-graph of a graph $\ccalG$ is the graph $\ccalG_\text{LG}$ whose nodes correspond to the edges of $\ccalG$.
Two nodes in $\ccalG_\text{LG}$ are connected if the corresponding edges in the original graph $\ccalG$ share an incident node.
Given the incidence matrix $\bbB$ of the original graph, the adjacency matrix of the line-graph can be represented as $\bbA_\text{LG} := |\bbB^\top\bbB - 2\bbI|$, where the absolute value is applied elementwise.
The graph Laplacian of the line-graph is correspondingly defined as $\bbL_\text{LG} = \diag(\bbA_\text{LG}\bbone) - \bbA_\text{LG} = \bbB_\text{LG}^{}\bbB_\text{LG}^\top$, where $\bbB_\text{LG}$ is the incidence matrix of the line-graph.

\section{Edge-space filtering -- an illustrative example}\label{S:motivation_contribution}
To understand the need for a theory of GSP for signals defined on the edges of a graph, let us consider the example graph in Fig.~\ref{fig:schematic_example}.
We assume that the (a priori undirected) graph has been endowed with oriented flows $\bbf_0$ on the edges as displayed in Fig.~\ref{fig:schematic_example}A.
However, we only get to observe a perturbed version $\bbf = \bbf_0 + \bbeps$ that has been distorted by white Gaussian noise $\bbeps$ (Fig.~\ref{fig:schematic_example}B).
Our goal is now to \emph{smooth} this observed signal $\bbf$ to get a better estimate of the original flows $\bbf_0$.

Translating the techniques previewed in \eqref{E:node_denoising} and \eqref{E:node_smoothing} into edge-space analysis, we may apply these procedures by treating the observed edge-flows $\bbf$ as node-data on the line-graph $\ccalG_\text{LG}$ associated with the original network. 
Fig.~\ref{fig:schematic_example}C shows the result of applying the line-graph filtering as in \eqref{E:node_smoothing} substituting $\bbL_\text{LG}$ for $\bbL$, and setting $k=10$ and $\mu=1/5$. 
The differences between adjacent edge signals are smoothed-out, and the output signal $\hat{\bbf}_\text{LG}$ is driven towards the global average of the input flows $\bbf$.
Indeed, we may interpret each smoothing operation $(\bbI - \mu \bbL_\text{LG})$ as a gradient descent step for the objective $\min_\bbf \|\bbB_\text{LG}^\top\bbf\|^2$, where each iteration drives down the difference between adjacent edge flows. 
For a connected line-graph, this (local) averaging behavior leads ultimately towards a global averaging of the flows, akin to a consensus protocol.

Although the above discussion provides a good understanding of the line-graph filtering operation, the filtering results are clearly not satisfactory for our considered example: the structure of the initial flows $\bbf_0$ has been distorted, and the error between the noise-free and the filtered signal $\|\bbf_0 - \hat{\bbf}_\text{LG}\|_2 \approx 6.68$ has in fact \emph{increased} when compared to the unfiltered signal $\|\bbf_0 - \bbf\|_2 \approx 3.83$.
The problem here stems from the notion of smoothness that we inherited from the node-based filter, where low-pass signals are associated with the small eigenvalues of the graph Laplacian. 
This assumption is well matched to applications in which we expect small variation of the signal across connected nodes, e.g., if the graph corresponds to a network of sensors that aims to monitor a smoothly varying scalar field such as a temperature.
However, in the case of flow data considered here this notion of a smooth signal is not desirable.
Arguably, a notion of smoothness for flow data should capture the idea of flow conservation, i.e., input and output at every node should (approximately) balance out.
Based on this view, cyclic flows such as depicted in Fig.~\ref{fig:schematic_example}A would correspond to smooth flows.

Interestingly, precisely this notion of smooth flows can be captured by the so called Edge-Laplacian defined as $\bbL_1 := \bbB^\top \bbB^{}$.
The Edge-Laplacian, which may be interpreted as a special case of the more general Hodge-Laplacian, has many properties that render this operator especially useful for flow-filtering, as we discuss in the next section.

\section{Edge-space filtering -- theory}\label{S:Hodge_filters}

The $\bbL_1$ Edge-Laplacian as an object on its own interest has been studied so far only sparsely in the literature, e.g., in the context of consensus protocols~\cite{zelazo_edge_laplacian_2011}, or as a special case of the more general Hodge-Laplacian (see e.g.~\cite{horak_edge_laplacian_2013,Lim2015,Schaub2018}). 
In the context of GSP, it has been recently considered in an independent work~\cite{barbarossa_higher_2018}.
We now present a number of important properties of this linear operator.
\begin{myproposition}\label{P:spectral_prop}
    Given a graph with Edge-Laplacian $\bbL_1 = \bbB^\top \bbB$, the following properties hold:
\begin{enumerate}
    \item The null space of $\bbL_1$ is equivalent to the cycle space $\mathcal F_C= \mathrm{ker}(\bbB)$, also known as the space of harmonic or cyclic flows.
    \item The image of $\bbL_1$ is equivalent to the gradient space ${\mathcal F_G=\mathrm{im}(\bbB^\top)}$, also known as the cut space or space of potential flows.
\end{enumerate}
\end{myproposition}
\begin{myproof}
    To prove \emph{(1)}, note that if $\bbz \in \mathrm{ker}(\bbB)$ then $\bbz \in \mathrm{ker}(\bbL_1)$, from the definition of $\bbL_1$. 
	For the converse, assume that $\bbL_1 \bbz = 0$, then $\bbz^\top \bbL_1 \bbz = \| \bbB \bbz \|_2^2 = 0$, so that $\bbz \in \mathrm{ker}(\bbB)$ concluding the proof.
	
    For \emph{(2)}, we proceed similarly. Clearly, $\mathrm{im}(\bbL_1) \subset \mathrm{im}(\bbB^\top)$. Conversely, for every non-zero $\bbz \in \mathrm{im}(\bbB^\top)$ we can find some $\bby \perp \mathrm{ker}(\bbB^\top)$ such that $\bbz = \bbB^\top \bby \neq 0$. Since this implies that $\bby \in \mathrm{im}(\bbB)$, there exists some $\bbx$ such that $\bbz = \bbB^\top \bbB \bbx = \bbL_1 \bbx$, and hence $\mathrm{im}(\bbB^\top) \subset \mathrm{im}(\bbL_1)$. 
\end{myproof}

Note that the name cycle space for $\mathcal F_C$ derives from the fact that $\mathrm{ker}(\bbB)$ is the set of edge-vectors corresponding to (oriented) cycles on the graph, i.e. edge-vectors that fulfill a flow conservation  constraint at every node. It can further be shown that the dimension of $\mathcal F_C$ is equal to the number of independent cycles in the graph~\cite{godsil_algebraic_2001}.
The gradient space $\mathcal F_G$ derives its name from the fact that it corresponds exactly to all those flow vectors $\bbf_\phi = \bbB^\top \bbphi \in \mathbb{R}^E$ that can be induced as the difference of a scalar potential function $\bbphi \in \reals^N$ defined on the nodes.

Proposition~\ref{P:spectral_prop} hints at the fact that an ideal low-pass filter $\bbH_{\mathrm{LP}}$ defined on $\bbL_1$ should cancel out all non-cyclic flows. This intuition can be formalized as follows. 
Denote by $\lambda_i(\bbL_1)$ the $i$-th eigenvalue of the matrix $\bbL_1$, then defining the frequency response $\tbh_{\mathrm{LP}}$ of an ideal low-pass filter (cf.~Section~\ref{S:graphs_background}) as
\begin{equation}\label{E:freq_resp_low_pass}
[\tbh_{\mathrm{LP}}]_i = 
\begin{cases}
1 \qquad \text{if} \,\, \lambda_i(\bbL_1) = 0,\\
0 \qquad \text{otherwise},
\end{cases}
\end{equation}
the following result holds.

\begin{myproposition}\label{P:ortho_projection}
	Given any edge-signal $\bbf$, it holds that
	\begin{equation}\label{E:prop_filtering}
	\bbH_{\mathrm{LP}} \bbf = \argmin_{\hbf} \|\hbf - \bbf\|_2^2 \quad  \mathrm{s.t.} \,\,\, \bbB \hbf = 0.
	\end{equation}
\end{myproposition}
\begin{myproof}
	Consider the Lagrangian $\ccalL(\hbf, \bbnu) = \|\hbf - \bbf\|_2^2 + \bbnu^\top \bbB \hbf$ associated with problem \eqref{E:prop_filtering}, and denote by $\hbf^*$ and $\bbnu^*$ the optimal primal and dual solutions of the problem. From the conditions of optimality if follows that
	\begin{align}
	\frac{\partial \ccalL}{\partial \hbf} & = 2 (\hbf^* - \bbf) + \bbB^\top \bbnu^* = 0, \label{E:conditions_optimality_lagrangian_1} \\
	\frac{\partial \ccalL}{\partial \bbnu} & = \bbB \hbf^* = 0. \label{E:conditions_optimality_lagrangian_2}
	\end{align}
	Left-multiplying \eqref{E:conditions_optimality_lagrangian_1} by $\bbB$, and substituting \eqref{E:conditions_optimality_lagrangian_2} into the obtained expression, it follows that $\bbB \bbB^\top \bbnu^* = 2\bbB \bbf$. From here we see that an optimal dual solution for $\bbnu$ is given by the node potentials $\bbnu^* = 2\bbL^\dagger\bbB\bbf$.
	Plugging this equality back into \eqref{E:conditions_optimality_lagrangian_1}, and using the properties of the pseudoinverse it immediately follows that $\hbf^* = (\bbI - \bbB^\top\bbL^\dagger \bbB) \bbf =  ( \bbI - \bbB^\dagger\bbB) \bbf$, indicating that $\hbf^*$ is the orthogonal projection of $\bbf$ onto the kernel of $\bbB$.
	Now \eqref{E:prop_filtering} follows from Proposition~\ref{P:spectral_prop}.
\end{myproof}

From the frequency response in \eqref{E:freq_resp_low_pass} it follows that the filter $\bbH_{\mathrm{LP}}$ outputs a projection of the input onto the kernel space of $\bbL_1$. 
Proposition~\ref{P:ortho_projection} formalizes the action of this filter from a denoising viewpoint.
More precisely, the constraint $\bbB \hbf = 0$ guarantees that $\hbf$ represents a flow-preserving edge signal. Thus, in \eqref{E:prop_filtering} we are seeking for the closest signal to $\bbf$ that is flow preserving. Proposition~\ref{P:ortho_projection} shows that such a signal is obtained by filtering out all the non-cyclic components present in~$\bbf$.

\noindent\textbf{Low-pass filtering in the edge-space. } 
Proposition~\ref{P:ortho_projection} notwithstanding, in many scenarios ideal low-pass filters are either undesirable or untenable in practice. Thus, leveraging the result in Proposition~\ref{P:spectral_prop}, we can define the analogues to \eqref{E:node_denoising} and \eqref{E:node_smoothing} for filtering flow signals $\bbf$ in the edges. More precisely, we have
\begin{align}
\hat{\bbf} = (\bbI + \alpha\bbL_1)^{-1}\bbf \qquad \text{(flow denoising)}, \label{E:flow_denoising}
\end{align}
where \eqref{E:flow_denoising} can be deemed as the solution of a regularized least-squares recovery akin to \eqref{E:regularized_least_squares}, and
\begin{align}
\hat{\bbf} = (\bbI - \mu \bbL_1)^k \bbf \qquad \text{(flow smoothing)}, \label{E:flow_smoothing}
\end{align}
which can be applied iteratively via $k$ successive applications of a single smoothing operator $(\bbI - \mu \bbL_1)$.

In order to illustrate the above filtering operations, let us apply \eqref{E:flow_smoothing} with $\mu=1/5$ and $k=10$ to 
the noisy flow in Fig.~\ref{fig:schematic_example}B, leading to the result in Fig.~\ref{fig:schematic_example}D.
As we can see, in this case the filtering leads to a much more desirable outcome: the cyclic structure of the original flow is essentially recovered, and the signal error $\|\bbf_0 - \hat{\bbf}_{\bbL_1}\|_2$ is reduced when compared to the original signal.
Formally, from Proposition~\ref{P:spectral_prop} we can conclude that the filtering operations \eqref{E:flow_denoising} and \eqref{E:flow_smoothing} drive the input towards the cycle space $\mathcal F_C$.
To illustrate this, we further plot the decomposition of the observed flows $\bbf$ onto the cycle (harmonic) and the gradient spaces in Fig.~\ref{fig:schematic_example}E.
Crucially, the cycle space is here of dimension $2$, meaning that the filter accounts for the two `degrees of freedom' in the original flow profiles (given by the two cycles), which explains the success of the filter in approximating the original flows.
In contrast, the null-space of the line-graph Laplacian $\bbL_\text{LG}$ is only one-dimensional, meaning that the filtered signal is driven towards a more restricted space, which is here completely unrelated to the cycle space.

The flow denoising procedure in \eqref{E:flow_denoising} can be deemed as the solution of a specific instance of a regularized least-squares problem where there is no potential function $\bbphi$ defined on the nodes, i.e., where the regularizer promotes flow conservation. 
In general, one might know the location of sources and sinks within the network, which can then be incorporated into the constraints in a formulation like~\eqref{E:prop_filtering}, or included as a modified regularizer.
This motivates the following optimization problem given a noisy flow signal $\bbf$
$\min_\hbf\lbrace \|\hbf - \bbf\|_2^2 + \alpha \|\bbB \hbf -  \bbphi \|_2^2\rbrace$,
where the denoised signal $\hbf$ is sought to be close to the observed $\bbf$ while approximately satisfying the flow balance equation. 
The solution to this augmented problem is
$\hat{\bbf} = (\bbI + \alpha\bbL_1)^{-1} (\bbf + \alpha \bbB^\top \bbphi)$.
Notice that, as expected, this reduces to \eqref{E:flow_denoising} in the absence of sources and sinks ($\bbphi = \mathbf{0}$).

\noindent\textbf{Mixed edge filters.}
We have thus far unveiled that the processing of signals on the edges of a graph is substantially different when considering the Laplacian of the line-graph $\bbL_\text{LG}$ as opposed to the Edge-Laplacian $\bbL_1$. More precisely, the latter is preferable when considering flow signals that are subject to (approximate) conservation laws. 
However, promoting similar signal values among incident edges -- as done by low-pass filters based on $\bbL_\text{LG}$ -- can be useful in different settings. 
For example, consider a bipartite graph representing students and homework problems, where the edges encode the perceived difficulty of each problem. Most probably some problems will be deemed as hard (or easy) by most of the students, thus presenting a smooth variation in the line graph.
In general, \emph{mixed edge filters} can combine the processing capabilities of $\bbL_\text{LG}$ and $\bbL_1$. More precisely, given a noise signal $\bbf$ on the edges of a graph, we define the filtered output as
\begin{equation}\label{E:line_edge_filter}
\hat\bbf_\text{mixed} = (\bbI + \alpha \bbL_1 + \beta \bbL_\text{LG})^{-1} \bbf,
\end{equation}
which is the optimal solution of the doubly-regularized problem
$\min_{\hbf} \lbrace \|\hbf - \bbf\|_2^2 + \alpha \hbf^\top \bbL_1 \hbf + \beta \hbf^\top \bbL_\text{LG} \hbf\rbrace$. Mixed edge filters as in~\eqref{E:line_edge_filter} can be especially relevant for the \emph{interpolation} of flow signals where large portions of the edges are unobserved, thus, flow conservation alone is not a good-enough regularizer. 
However, a detailed treatment of the problem of interpolation of edge-signals is left as future work.

\section{Numerical experiments}\label{S:Simulations}
\begin{figure}[tb!]
    \centering
    \includegraphics[width=\linewidth]{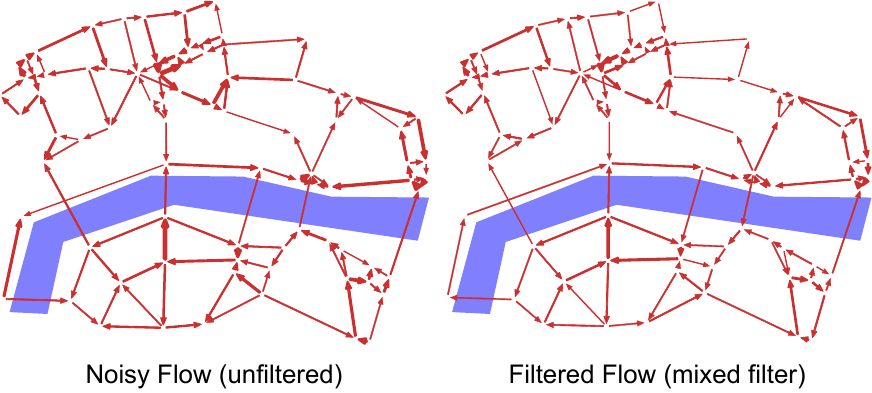}
    \caption{\textbf{Flow denoising on the London street network.}
    Left: unfiltered noisy flow on the London street network. Right: filtered flows using the mixed flow-denoising discussed method in \eqref{E:line_edge_filter} with parameters $\alpha=28, \beta=0.06$, chosen via a grid-search. Note how some flows have switched direction after filtering.}
\label{fig:london_streets}
\end{figure}

We consider a (subset) of the street network of London -- as considered in~\cite{Youn2008,schaub_network_traffic_2014} -- that we illustrate in Fig.~\ref{fig:london_streets} along with the River Thames for geographical orientation.
The network comprises 82 crossings (nodes) and 130 streets (edges) on which we place a generic initial flow $\bbf_0$ distorted by white Gaussian noise (see Fig.~\ref{fig:london_streets}-left).
Since most real edge data is neither perfectly flow-preserving nor exactly constant, the initial flow $\bbf_0$ was generated as a random mixture of a harmonic flow and a potential flow based on the line-graph Laplacian, thus containing low-pass components according to both filter classes.

On this data we apply a line-graph denoising (with parameter $\alpha=0.16$), the flow denoising procedure~\eqref{E:flow_denoising} ($\alpha=37$), and the mixed denoising filter~\eqref{E:line_edge_filter} ($\alpha = 28,\beta=0.06$), where all the parameters have been chosen through a grid search for best performance.
In this scenario, the initial error $\|\bbf_0 - \bbf\|_2 \approx 8.45$ is reduced by all the filters.
The line-graph filter leads to an error of $\|\bbf_0 - \hat\bbf_\text{LG}\|_2 \approx 7.08$.
The Edge-Laplacian filter provides a better result with $\|\bbf_0 - \hat\bbf_{\bbL_1}\|_2 \approx 5.65$.
However, the best results are achieved by the mixed edge filter with an error $\|\bbf_0 - \hat\bbf_\text{mixed}\|_2 \approx 5.24$, depicted in Fig.~\ref{fig:london_streets}-right. The superior performance of the mixed edge filter indicates that both the conventional line-graph filters and the proposed Edge-Laplacian filters are indeed well suited to eliminate different types of noise.
A more detailed examination of real flow data, and a characterization of the settings where mixed filters achieve marked performance improvements -- in terms of the noiseless signals, the type of noise present, and the underlying graph topology -- are left as future work.

\section{Conclusions}\label{S:Conclusions}

We presented a GSP framework for the treatment of signals defined on the edges of a graph. 
Instead of considering the line-graph and replicating the tools of node-centric GSP, we proposed a different approach based on the Edge-Laplacian. We showed that the proposed approach is indeed better suited for the treatment of signals representing flows in networks, and presented a series of filters for the denoising and smoothing of flow signals. Finally, we illustrated that the line-graph and Edge-Laplacian approaches can be combined, obtaining better denoising results in practice.

A number of exciting future directions follow naturally from the work here presented including: i) Edge-Laplacian methods for interpolation and treatment of missing data; ii) Non-linear filtering in the edge-space; and iii) Extensions to signals defined on higher-order simplicial complexes.

\bibliographystyle{IEEEbib}
\bibliography{citations}

\end{document}